\documentclass[nonacm,sigplan,screen]{acmart}
\usepackage{pervasives}

\usepackage[normalem]{ulem}

\author{Caleb Kim}
\authornote{Equally contributing authors.}
\affiliation{
  \institution{Cornell University}
  \country{USA}
}
\author{Pai Li}
\authornotemark[1]
\affiliation{
  \institution{Cornell University}
  \country{USA}
}
\author{Anshuman Mohan}
\affiliation{
  \institution{Cornell University}
  \country{USA}
}
\author{Andrew Butt}
\affiliation{
  \institution{Cornell University}
  \country{USA}
}
\author{Adrian Sampson}
\affiliation{
  \institution{Cornell University}
  \country{USA}
}
\author{Rachit Nigam}
\affiliation{
  \institution{Cornell University}
  \country{USA}
}

\begin{document}

\title{Unifying Static and Dynamic Intermediate Languages for Accelerator Generators}

\begin{abstract}
Compilers for accelerator design languages (ADLs) translate high-level languages into application-specific hardware.
ADL compilers rely on a hardware \emph{control interface} to compose hardware units.
There are two choices:
\emph{static} control,
which relies on cycle-level timing;
or \emph{dynamic} control,
which uses explicit signalling to avoid depending on timing details.
Static control is efficient but brittle;
dynamic control incurs hardware costs to support compositional reasoning.

\sys{} is an ADL compiler that unifies static and dynamic control in a single intermediate language (IL).
Its key insight is that the IL's static fragment is a \emph{refinement} of its dynamic fragment:
static code admits a subset of the run-time behaviors of the dynamic equivalent.
\sys{} can optimize code by combining facts from static and dynamic submodules,
and it opportunistically converts code from dynamic to static control styles.
%
We implement \sys{} as an extension to an existing dynamic ADL compiler, Calyx.
We use \sys{} to implement an MLIR frontend,
a systolic array generator,
and a packet-scheduling hardware generator
to demonstrate its optimizations and the static--dynamic interactions it enables.
\end{abstract}

\maketitle

\section{Introduction}

Accelerator design languages (ADLs)~\cite{dahlia,spatial,aetherling,darkroom,polysa} raise the level of abstraction for hardware design.
The idea is analogous to traditional software compilation:
we want users to work not with gates, wires, and clock cycles, but with high-level or domain-specific concepts such as tensor operations~\cite{heterocl}, functional programs~\cite{aetherling,halide-hls}, and recurrence equations~\cite{polysa}.
Compilers then translate these high-level descriptions into efficient hardware designs.
ADLs suffer cross-cutting compilation challenges, and the architecture community has responded with a range of compiler frameworks and intermediate languages~\cite{hir, muir, hector, circt-latte}.

This paper identifies a central challenge for ADL compilers: the \emph{control interface} for composing units of hardware.
The choice of interface has wide-ranging implications on a compiler's expressive power, its ability to optimize programs, and the semantics of its intermediate language.
There are two categories.
\emph{Dynamic} or \emph{latency-insensitive} interfaces abstract away timing details and streamline compositional design, but they incur fundamental overheads~\cite{licost}.
\emph{Static} or \emph{latency-sensitive} interfaces are efficient, but they depend on the cycle-level timing of each module and therefore leak implementation details across module boundaries.

Intermediate languages (ILs) for ADLs use either
dynamic interfaces~\cite{calyx,dynamic-schedule-hls},
static interfaces~\cite{hir}, or both~\cite{muir,hector}.
Static interfaces alone are insufficient because some computations, such as off-chip memory accesses, have fundamentally variable latencies.
Infrastructures that support both interfaces typically \emph{stratify} the IL into separate dynamic and static sub-languages~\cite{dass,hector}.
While stratified compilers can bring customized lowering and optimization strategies to bear on each sub-language,
they entail duplicated implementation effort and miss out on cross-cutting optimizations that span the boundary between static and dynamic code.
Stratification also infects the frontends targeting the IL: they must carefully separate code between the two worlds and manage their interaction.

We introduce \sys{}, an IL and compiler for accelerator designs that freely mix static and dynamic interfaces.
The key insight is that static IL constructs are all \emph{refinements} of their dynamic counterparts:
they admit a subset of the run-time behaviors.
This unified approach lets transformations and optimizations work across both interface styles.
\sys{} also enables the incremental adoption of static interfaces:
frontends can first establish correctness using compositional but slow dynamic code,
and then opportunistically convert to efficient static interfaces.
Refinement in \sys{} guarantees that this transition is correct.

We implement \sys{} as an extension to the dynamic-first Calyx infrastructure~\cite{calyx}.
This paper shows how to compile \sys{}'s static extensions into pure Calyx.
We lift Calyx's existing optimizations to support \sys{}'s static abstractions and implement new time-sensitive optimizations.
\sys{} can also automatically infer when some dynamic Calyx code has fixed latency, and promote it to static code.

We evaluate \sys{}'s new optimizations
using a frontend that translates from high-level MLIR~\cite{mlir} dialects to \sys{}.
Time-sensitive optimizations improve execution times by $2.5\times$ on average over dynamic Calyx code.
We also implement a packet-scheduling engine and study how \sys{} optimizations, in concert with domain-specific human insight, are able to improve the performance of the generated hardware.
As another domain-specific case study, we extend a systolic array generator to support fused dynamic operations to understand how \sys{} can support interactions between fundamentally static and dynamic components. \xxx[C]{Do we want to briefly mention performance improvements too?}

\section{Hardware Interfaces}
\label{section:motivation}

\begin{figure}
  \centering
  \begin{subfigure}[t]{\linewidth}
    \centering
    \includegraphics[scale=0.9]{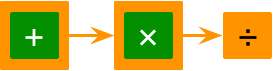}
    \vspace{-5pt}
    \caption{A dynamic IL wraps static components with dynamic wrappers.}
    \label{fig:evo:dynamic}
  \end{subfigure}
  \begin{subfigure}[t]{\linewidth}
    \centering
    \includegraphics[scale=0.9]{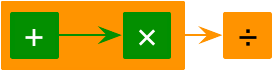}
    \vspace{-5pt}
    \caption{A \emph{stratified} IL uses one compilation path for static ``islands'' and a separate one that incorporates dynamic components.}
    \label{fig:evo:stratified}
  \end{subfigure}
  \begin{subfigure}[t]{\linewidth}
    \centering
    \includegraphics[scale=0.9]{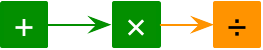}
    \vspace{-5pt}
    \caption{\sys{} represents static dynamic and static components in the same IL. The compiler can perform whole-program optimizations.}\label{fig:evo:calyx}
  \end{subfigure}
  \vspace{-5pt}
\caption{Hardware implementations of $(a+b)\times c \div d$. Green units have static interfaces; orange is dynamic.}
    \label{fig:evo}
\end{figure}

Consider compiling the integer computation ${(a+b)\times c \div d}$ into hardware.
Generating a hardware datapath entails orchestrating physical units---such as adders, multipliers, and dividers---over time.
For this example, we use sequential, i.e., non-pipelined, hardware units.
%
%
While many hardware units, such as adders and multipliers, have fixed latency, many do not:
integer dividers, for instance,
typically have data-dependent timing.
Control logic for these two categories is fundamentally different.
A variable-latency divider may expose 1-bit signal wires to start computation and to signal completion.
A fixed-latency multiplier, however, needs no explicit completion signalling:
clients can simply provide inputs and wait the requisite number of cycles.
%
%
For this example, assume we have an adder with latency~1, a multiplier with latency 3, and a divider with variable latency.

\paragraph{Dynamic compilation.}
\Cref{fig:evo:dynamic} shows how dynamic-first ILs, such Calyx~\cite{calyx} and Dynamatic~\cite{dynamic-schedule-hls}, might compile our expression.
All units expose explicitly signalled dynamic interfaces;
each static module requires a \emph{wrapper} that
counts clock cycles up to the unit's latency and then signals completion.
A purely dynamic compiler benefits from a uniform interface and
compositional reasoning,
because no module can depend on the timing of any other.
However, these wrappers incur time and space overheads,
and optimizations cannot exploit timing information
(\cref{sec:optimizations:compaction}).

\paragraph{Static compilation.}
Static-first ILs,
such as HIR~\cite{hir}, require fixed-latency operations.
They can support dynamic operators like dividers by using an upper-bound latency.
Upper bounds are pessimistic, however,
and some hardware operations have unbounded latency:
the latency for an arbiter that manages conflicting memory accesses, for instance, fundamentally depends on the address stream.

\paragraph{Stratified static--dynamic compilation.}
\Cref{fig:evo:stratified} illustrates a hybrid approach,
such as DASS~\cite{dass}, that combines static and dynamic compilation.
The idea is to compile the two parts of the program separately:
first using static interfaces for the fixed-latency fragment, $(a + b) \times c$, and then using dynamic interfaces to combine this fragment with the variable-latency divider.
This combination allows latency-sensitive optimizations on the static fragment while still allowing dynamic scheduling where it is beneficial.

This \emph{stratified} approach, however, needs separate ILs for the two styles of computation.
The compiler cannot exploit information across the static--dynamic boundary.
Furthermore, it complicates the job for frontends that emit these ILs:
switching a single subcomputation from static to dynamic requires a global change in the way the program is encoded.

\paragraph{Unified static--dynamic compilation in \sys{}.}
\Cref{fig:evo:calyx} represents our approach with \sys{}: a unified IL that expresses both static and dynamic interfaces in one program.
\sys{} extends Calyx~\cite{calyx}, an existing dynamic-first IL, with static constructs that \emph{refine} the semantics of its dynamic constructs.
By mirroring the dynamic IL abstractions with static counterparts, \sys{}
enables compositional reasoning, incremental adoption, and whole-program optimization across the static--dynamic boundary.

\begin{figure}
\begin{lstlisting}[language=calyxspec,numbers=left, belowskip=-0.8\baselineskip]
component expr(a:32,b:32,c:32,d:32)->(out:32) {
  cells {
    add = std_add(32);   // 32-bit adder
    mult = std_mult(32); // 32-bit multiplier
    div = std_div(32);   // 32-bit divider
  }
  wires {
    @static<1>@ group do_add {
      add.left = @%[0:1]@ ? a;
      add.right = @%[0:1]@ ? b;
      (*{\color{red}//}*) do_add[done] = add.done;
    }
    @static<3>@ group do_mult {
      mult.left = @%[0:3]@ ? add.out;
      mult.right = c; // implicit %[0:3] guard
      (*{\color{red}//}*) do_mult[done] = mult.done;
    }
    group do_div {
      div.go = 1'd1;
      div.left = mult.out;
      div.right = d;
      do_div[done] = div.done;
    }
    out = div.out;
  }
  control {
    seq { @static@ seq { do_add; do_mult; }
          do_div;
    }
  }
}
\end{lstlisting}
\caption{A \sys{} component that computes $(a + b) \times c \div d$. Our extensions to the language are shown in red.}\label{fig:calyx-expr}
\end{figure}

\section{The \sys{} Intermediate Language}\label{sec:calyx}
\label{section:calyx}

This section introduces \sys{}, a \emph{unified} IL for compiling hardware accelerators.
\sys{} extends Calyx~\cite{calyx}, an existing dynamic IL.
Calyx has a growing family of frontends,
such as
for Halide~\cite{ragan-kelley:halide, halide-to-calyx}
and MLIR dialects in CIRCT~\cite{circt-latte, zang-sycl-circt},
that can adopt \sys{}'s static interfaces
to improve performance.

We introduce \sys{} using the program in
\cref{fig:calyx-expr} as a running example.
Our extensions to Calyx are in red.
Deletions when porting from Calyx to \sys{} are commented out with red slashes.
We describe the existing Calyx IL (\cref{sec:classic-calyx}),
show that its original \emph{hint-based} treatment of static interfaces is insufficient
(\cref{sec:hints-not-enough}),
and then introduce \sys{}'s extensions.

\subsection{The Calyx IL}
\label{sec:classic-calyx}

The Calyx IL intermixes software-like \emph{control operators} with hardware-like \emph{structural resources}~\cite{calyx}.
The former simplifies encoding of high-level language abstractions,
while the latter enables optimizations that exploit control information to optimize the physical hardware implementation.


\paragraph{Components.}
Components define units of hardware with input and output \emph{ports}.
In \Cref{fig:calyx-expr}, \code{expr} has four $32$-bit input ports (\code{a} through \code{d}) and one output port (\code{out}).
A component has three sections: \code|cells|, \code|wires|, and \code|control|.

\paragraph{Cells.}
The \code|cells| section instantiates subcomponents.
Cells can be either other Calyx components or \emph{external definitions} defined in a standard HDL.
The component \code|expr| instantiates three cells from the standard library:
\code|add|,
\code|mult|, and
\code|div|.
Each is parameterized by a bitwidth.


\paragraph{Wires.}
Calyx uses \emph{guarded assignments} to connect two ports when a logical condition, called the \emph{guard}, is true.
Consider:
%
\begin{lstlisting}
add.left = c0 ? 10;
add.left = c1 ? 20;
add.right = 30;
\end{lstlisting}
Here, \code|add.left| has the value $10$ or $20$ depending on which guard, \code|c0| or \code|c1|, is true.
Meanwhile, \code|add.right| \emph{unconditionally} has the value $30$.
Calyx's well-formedness constraint requires that all guards for a given port be mutually exclusive:
it is illegal for \code|c0| and \code|c1| to simultaneously be true.

\paragraph{Groups.}
Assignments can be organized into unordered sets called \emph{groups}.
A group can execute over an arbitrary number of cycles and therefore requires a $1$-bit \code|done| condition to signal completion.
In \cref{fig:calyx-expr}, the assignments in \code|do_div| compute \code{mult.out} $\div$ \code{d} by passing in inputs and asserting the divider's ``start'' signal, \code|div.go|.
The group's \code|done| signal is connected to the divider's \code|done| port,
which becomes~1 when the divider finishes.

\paragraph{Control.}
The control section is an imperative program that decides when to execute groups.
Calyx supports sequential (\code{seq}), parallel (\code{par}), conditional (\code{if}), and iterative (\code{while})
 composition.
The \code{if} and \code{while} constructs use one-bit condition ports.
An \code|invoke| operator is analogous to a function call:
it executes the control program of a subcomponent fully and then returns control to the caller.

\subsection{Latency Sensitivity in Calyx}
\label{sec:hints-not-enough}

As a fundamentally dynamic language,
Calyx \emph{provides no guarantees on inter-group timing} in its control programs:
programs cannot rely on the relative execution schedule of any two groups.
For example, any amount of time may pass between steps in a \code|seq| block; and different threads in a \code|par| block may start at different times, so no thread may rely on the timing of another~\cite{cidr}.
The compiler exploits this semantic flexibility to optimize programs
by adjusting this timing.

However, latency insensitivity is expensive~\cite{licost}.
To help mitigate this cost, Calyx comes with an optional attribute, \code|@static($n$)|, that \emph{hints} to the compiler that a group or component has a fixed latency of $n$ cycles.
These hints do not affect the program's semantics, so the compiler can disregard or erase them.
However, this optional nature makes them
challenging to support and reason about.
Each compiler optimization pass must treat the hint \emph{pessimistically}; there is no contract to maintain time-sensitive behavior, which has lead to several bugs~\cite{calyx-par-bug-1,calyx-par-bug-2}.
Erasable hints are also unsuitable for integrating with external hardware,
such as a module that produces an answer exactly 4 cycles after reading an input.
This paper's thesis is that the distinction between static and dynamic control is too important---and too semantically meaningful---to be encoded as an optional hint.
Instead, the IL's static constructs must be a \emph{semantic refinement} (~\cref{sec:why-extend}) of its dynamic equivalents:
converting from dynamic to static restricts a program's timing behavior;
the reverse is not allowed because it allows more possible behaviors.



\subsection{Static Structural Abstractions}
\label{sec:calyx:static-groups}

\sys{} extends Calyx with new, time-sensitive structural abstractions: static components and static groups.

\paragraph{Static components.}
\sys{}'s static components are like Calyx's dynamic components, but they use a different ``calling convention.''
Where dynamic components, such as \code|std_div|, use a \code|go| signal to start computation and a \code|done| signal to indicate completion,
static components only use \code|go|.
Compare the interface of a multiplier to that of a divider:
\begin{lstlisting}[language=calyxspec]
@static<3>@ primitive std_mult[W](
  go: 1, left: W, right: W) -> (out: W);
primitive std_div[W](
  go: 1, left: W, right: W) -> (out: W, done: 1)
\end{lstlisting}
The \code|static<$n$>| qualifier indicates a latency of $n$ cycles
that is guaranteed to be preserved by the \sys{} compiler.

\paragraph{Static groups and relative timing guards.}
Static groups in \sys{}
use \emph{relative timing guards}, which allow assignments on specific clock cycles.
This group computes $\textit{ans} = 6 \times 7$:
\begin{lstlisting}[language=calyxspec,numbers=left, firstnumber=1]
static@<4>@ group mult_and_store {
  mult.left = @%[0:3]@ ? 6;
  mult.right = @%[0:3]@ ? 7;
  mult.go = @%[0:3]@ ? 1;    // run the multiplier
  ans.in = @%3@ ? mult.out;   // ans is a register
  ans.write_en = @%3@ ? 1;  // assert write enable
}
\end{lstlisting}
Like \code|do_div| in \cref{fig:calyx-expr}, the group sends operands into the \code|left| and \code|right| ports of an arithmetic unit.
Here, however, relative timing guards encode a cycle-accurate schedule:
a guard \code{
is true in the half-open interval from cycle~$i$ to cycle~$j$ of
the group's execution.
The assignments to ports \code|mult.left| and \code|mult.right| are active for the first~3 cycles.
The guard \code|
is syntactic sugar for \code{
so the write into the \code|ans| register occurs on cycle 3.
The \code|static<4>| annotation tells us the group is done
on cycle 4.

\sys{}'s relative timing guards resemble cycle-level schedules in some purely static languages~\cite{hir,filament}.
However, they count relative to the start of the \emph{group}, not that of the \emph{component}.
This distinction is crucial since it lets \sys{} use static groups in both static and dynamic contexts.

\subsection{Static Control Operators}\label{sec:calyx:static-control}

\sys{} provides a static alternative to each dynamic control operator in Calyx.
Unlike the dynamic versions, static operators guarantee specific cycle-level timing behavior.


The \code|static| qualifier marks static control operators.
While dynamic commands may contain both static and dynamic children, static commands must only have static children.
We write $|c|$ for the latency of a static command $c$.


\paragraph{Sequential composition.}
A \code{static seq} like this:
\begin{lstlisting}
static seq {$c_1$; $c_2$; ...; $c_n$;}
\end{lstlisting}
has a latency of $\sum_1^n |c_i|$ cycles.
$c_1$ executes in the interval $[0, |c_1|)$ after the \code|seq|'s start, $c_2$ in $[|c_1|, |c_1|+|c_2|)$, and so on.


\paragraph{Parallel composition.}
A \code{static par} statement:
\begin{lstlisting}
static par {$c_1$; $c_2$; ...; $c_n$;}
\end{lstlisting}
has latency
$\max_1^n |c_i|$.
Command $c_1$ is active between $[0, |c_1|)$, program $c_2$ between $[0, |c_2|)$, and so on.

The parallel threads in a \code|static par| can depend on the ``lockstep'' execution of all other threads.
Threads can therefore communicate, whereas conflicting parallel state accesses in Calyx are data races and therefore undefined behavior~\cite{cidr}.



\paragraph{Conditional.}
Static conditionals use a 1-bit port $p$:
%
\begin{lstlisting}
static if $p$ { $c_1$ } else { $c_2$ }
\end{lstlisting}
The latency is the upper bound of the branches, $\max(|c_1|, |c_2|)$.



\paragraph{Iteration.}
There is no static equivalent to Calyx's unbounded \code|while| loops.
\sys{} instead adds
both static and dynamic variants of fixed-bound \code|repeat| loops:
\begin{lstlisting}
static repeat $n$ { $c$ }
\end{lstlisting}
The body executes $n$ times, so the latency is $n \times |c|$.

\paragraph{Invocation.}
\sys{}'s \code|static invoke| corresponds to Calyx's function-call-like operation
and requires the target component to be static.
The latency is that of the invoked cell.

\paragraph{Group enable.}
A leaf statement can refer to a \code|static group| (e.g., \code|do_add| in \cref{fig:calyx-expr}).
The latency is that of the group.


\subsection{Unification Through Semantic Refinement}
\label{sec:why-extend}

\sys{}'s static constructs are all semantic \emph{refinements}~\cite{oprefinement} of their dynamic counterparts in Calyx.
The semantics of dynamic code admit many concrete execution schedules, such as arbitrary delays between group executions.
%
Each static construct instead selects one \emph{specific} cycle-level schedule from among those possibilities.
%

Refinement enables \emph{incremental adoption}:
a frontend can first generate purely dynamic code, establish correctness using the original Calyx semantics based on partial ordering between group executions, and then add \code|static| qualifiers.
We can establish correctness for the \code|static| code by the same argument as the original code, since it admits a subset of the original's cycle-level executions.
This implication also means that \sys{} may automatically infer \code|static| qualifiers for some code (\cref{sec:optimizations:compaction}).

Semantic refinement also enhances optimization (\cref{section:optimizations:cell-share}).
\sys{} can enrich existing Calyx passes with timing information to expose more optimization opportunities in static code.
New optimizations can also combine information across static and dynamic code.
This kind of optimization would
be challenging in a stratified compiler like DASS~\cite{dass} with separate ILs and lowering paths for static and dynamic code.

\section{Compilation}\label{sec:compilation}

\begin{figure*}
\centering
\includegraphics[width=\linewidth]{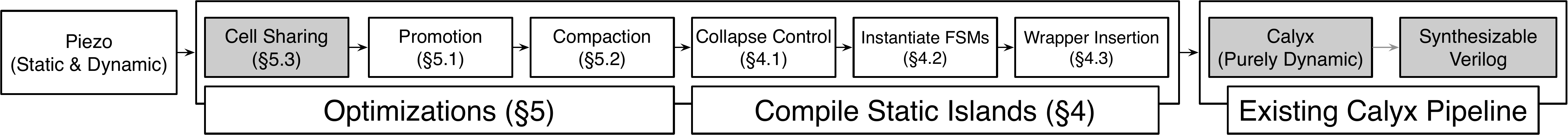}
\vspace{-20pt}
\caption{%
\sys{} compilation flow. The extended \sys{} syntax is optimized (\cref{sec:optimizations}) and compiled (\cref{sec:compilation}) to pure Calyx abstractions.}
\label{fig:comp-flow}
\end{figure*}

\begin{figure*}
\centering
\includegraphics[width=\linewidth]{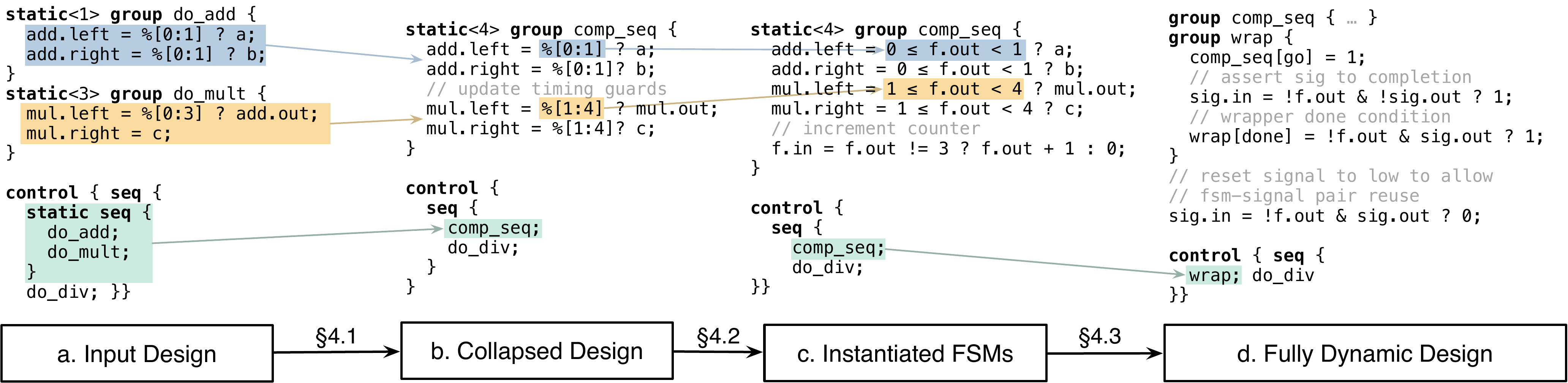}
\vspace{-20pt}
\caption{The stages for implementing \sys{}'s static control operators.
}
\label{fig:static-comp}
\end{figure*}

\begin{table}[]
\centering
\caption{Interfaces between types of control.}
\vspace{-10pt}
\begin{tabular}{llll}
\toprule
\textbf{Abbr.} & \textbf{Caller} & \textbf{Callee}  & \textbf{Calling Convention}  \\
\midrule
$D\rightarrow D$ & Dynamic & Dynamic & Calyx~\cite{calyx} \\
$S\rightarrow S$ & Static & Static  & See \cref{sec:compilation:collapse}          \\
$D\rightarrow S$ & Dynamic & Static  & See \cref{sec:compilation:groups}       \\
$S\rightarrow D$ & Static  & Dynamic  & Not supported  \\
\bottomrule
\end{tabular}
\label{tbl:calling-conventions}
\end{table}

\cref{fig:comp-flow} shows the compilation flow for \sys{}.
After optimizations (\cref{sec:optimizations}),
we translate \sys{} constructs to pure Calyx.
%
%
%
%

\label{sec:compilation:calling-convention}


The \sys{} compiler relies on control interfaces for static code, dynamic code, and invocations that cross the static--dynamic boundary.
For example, in a control statement like
\code|seq { a; b; }|,
both the parent (the \code{seq})
and the children (\code{a} and \code{b})
could use either static or dynamic control.

\Cref{tbl:calling-conventions}
lists the four possible cases,
denoted ${I_p \rightarrow I_c}$
where the parent and child interfaces $I$ are static ($S$) or dynamic ($D$).
The all-dynamic case,
${D \rightarrow D}$, is the Calyx baseline.
The all-static case, ${S \rightarrow S}$, works
by counting cycles (\cref{sec:compilation:collapse}).
For ${D \rightarrow S}$,
the compiler adds
a \emph{dynamic wrapper} around the static child (\cref{sec:compilation:groups}).
\sys{} disallows the ${S \rightarrow D}$ case with a compile-time error:
if the child takes an unknown amount of time, it is impossible to give the parent a static latency bound.
Given the prohibition against ${S \rightarrow D}$ composition,
we can think of any \sys{} program as a dynamic control program with interspersed \emph{static islands}~\cite{dass-islands,petri-islands}.

Compilation starts by \emph{collapsing} static islands into static groups (\cref{sec:compilation:collapse})
and then generating FSM logic to implement relative timing guards (\cref{sec:compilation:fsm-inst}).
Finally, it \emph{wraps} static islands for use in their dynamic context (\cref{sec:compilation:groups}).

\subsection{Collapsing Control}
\label{sec:compilation:collapse}


Figures~\ref{fig:static-comp}a--b illustrate how \sys{} \emph{collapses} static control statements into static groups.
The new group contains all the assignments from the old groups used in the statement (\code|do_add| and \code|do_mult| in the example),
with their timing guards updated to implement the statement's timing.


We collapse each static island in a \emph{bottom-up} order:
to compile any statement, we first collapse all its children.
Before collapsing,
we preprocess assignments to add timing guards where they are missing:
for example, the assignment \code|mul.right = c| in \cref{fig:static-comp}a is normalized to \code|mul.right = 
We combine these timing guards with any existing guards using the conjunctive operator \code|&|.


\paragraph{Parallel composition.}
With all timing guards explicit and the children already collapsed, compiling \code{static par} is simple:
we merge the assignments from the children into a single static group. The new group's latency is the maximum latency among the children.

For example, we compile:
\begin{lstlisting}
static<1> group A { r1.in = 1; r1.write_en = 1; }
static<2> group B { r2.in = 4; r2.write_en = 1; }
control { static par { A; B; } }
\end{lstlisting}
into:
\begin{lstlisting}
static<2> group comp_par {
  r1.in = %[0:1] ? 1; r1.write_en = %[0:1] ? 1;
  r2.in = %[0:2] ? 4; r2.write_en = %[0:2] ? 1;
}
control { comp_par; }
\end{lstlisting}

\paragraph{Sequential composition.}
To compile \code{static seq}, we can merge assignments from child groups.
We rewrite each timing guard \code|
%
The new group's latency is the sum of latencies of the children.
%
For example:
\begin{lstlisting}
control { static seq { A; B; } }
\end{lstlisting}
compiles
(where \code|A| and \code|B| are as above)
into:
\begin{lstlisting}
static<3> comp_seq {
  r1.in = %[0:1] ? 1; r1.write_en = %[0:1] ? 1;
  r2.in = %[1:3] ? 4; r2.write_en = %[1:3] ? 1;
}
control { comp_seq; }
\end{lstlisting}

\paragraph{Conditional.}
Semantically, \code{static if} only checks its condition port once:
it must ignore any changes to the port while either branch executes.
%
To honor this while compiling
\code|static if cond { A } else { B } }|,
we stash
\code|cond|'s value in a special register on the first cycle, and leave the register's value unchanged thereafter.
We generate logic to select between \code|A| and \code|B| using \code|cond| directly during the first cycle, and the special register for the remaining cycles.

\paragraph{Iteration.}
To implement \code|static repeat $n$ { $g$ }|,
the collapsed body group $g$ must run $n$ times.
Activating a static group in \sys{} entails asserting its \code|go| signal for the group's entire latency.
We can therefore compile the loop into a group that asserts $g$'s \code|go| signal for $n \times |g|$ cycles:
\begin{lstlisting}
static<$n \times |g|$> repeat_group { $g$[go] = 1; }
\end{lstlisting}
In this case, the body group $g$ remains alongside the new \code|repeat_group|.
The body group's FSM (see ~\cref{sec:compilation:fsm-inst}) is responsible for resetting itself every $|g|$ cycles.

%
%

\subsection{FSM Instantiation}
\label{sec:compilation:fsm-inst}

Figures~\ref{fig:static-comp}b--c illustrate the next compilation step:
eliminating static timing guards (\cref{sec:calyx:static-groups}).
For a static group with latency $n$,
this pass generates a finite state machine (FSM) counter
that counts from $0$ to $n-1$; it automatically resets back to $0$ immediately after hitting $n-1$.
%
We translate each timing guard \code{
into the guard $j \leq f < k$ where $f$ is the counter.

Resetting the counter from $n-1$ to $0$ lets
static groups re-execute immediately after finishing.
Compiled \code|repeat| and \code|while| loops, for example, can chain invocations of static bodies without wasting a cycle between each iteration.

While FSM instantiation would work the same on the original program, it is more efficient to run it after collapsing control.
Generating fewer static groups yields fewer FSM registers and incrementers.

\subsection{Wrapper Insertion}\label{sec:compilation:groups}

Figures~\ref{fig:static-comp}c--d illustrate the final compilation step:
converting each collapsed, timing-guard-free static group (\ref{fig:static-comp}c) into a dynamic group (\ref{fig:static-comp}d).
%

We generate a \emph{dynamic wrapper} group for every static group that has a dynamic parent.
Like any dynamic group,
the wrapper exposes two 1-bit signals, \code|go| and \code|done|.
%
%
When activated with \code|go|,
the wrapper in turn actives the \code|go| signal of the static group.
To generate the \code|done| signal, the wrapper uses a 1-bit signal \code{sig} to detect if a static island's FSM has run once.
When the FSM is 0 \emph{and} \code{sig} is high, we know that the FSM has \emph{reset} back to 0: the wrapper asserts \code|done|.

\paragraph{Special case: \code{while} with static body.}
The wrapper strategy works in the general case, but when the dynamic parent is a \code|while| loop,
the compiled code ``wastes'' one cycle per iteration to check the loop condition.
This strategy incurs a relative overhead of $1/b$ when the body takes $b$ cycles, which is bad for short bodies and large trip counts.
This special case is common because it lets programs build long-running computations from compact hardware operations, so we handle it differently to eliminate the overhead.

To compile \code|while $c$ { $g$ }| where $g$ is static,
we generate a wrapper for the entire \code|while| loop instead of a wrapper for~$g$ alone.
Each time the FSM returns to the initial state, the wrapper concurrently checks the condition port and asserts \code|done| if the condition is false.
This is another application of refinement in \sys{}:
Calyx's \code|while| operator admits multiple possible cycle-level timing behaviors, and we generate a specific one to meet our objectives.

\section{Optimizations}
\label{sec:optimizations}

We design a pass to opportunistically convert dynamic code to static code
and new time-sensitive static optimizations.


\subsection{Static Inference and Promotion}
\label{sec:optimizations:promotion}

Calyx code written as dynamic often does not need to be dynamic:
its latency is deterministic.
\emph{Promoting} such code to use static interfaces can save time and resources for dynamic signalling---but it is not always profitable.
We therefore split the process into two steps: \emph{inference}, which detects when dynamic groups and control have a static latency, and \emph{promotion},
which converts dynamic code to static code when it appears profitable.
Inference records information without affecting semantics, while promotion refines the program's semantics.
We infer freely but promote cautiously.


\paragraph{Inferring static latencies.}


We use an existing Calyx pass called \code|infer-static-timing| pass to infer latencies for both groups and control programs.
It infers a group's latency by analyzing its uses of its \code{go} and \code{done}.
Suppose we have:
%
\begin{lstlisting}[language=calyx]
group g {
  reg.in = 10;  // reg is a register (latency 1)
  reg.write_en = 1;
  g[done] = reg.done;  }
\end{lstlisting}
The pass observes that
(1) \code|reg.write_en| is asserted unconditionally,
(2) the group's \code|done| flag is tied to \code|reg.done|, and
(3) the register component definition declares a latency of 1.
Calyx therefore attaches a \code{@static(1)} annotation to~\code{g}: the group will take exactly one cycle to run.

For control operators, e.g., \code|seq|, inference works bottom-up.
If all of a \code|seq|'s children have \code{@static} annotations, the \code|seq| gets a \code{@static($n$)} annotation
where $n$ is the sum of the latencies of its children.
Despite this inference, Calyx's original
time-sensitive FSM generation pass cannot compile static control islands; instead, the entire component needs to be static~\cite{calyx-static-bug}.
\sys{} lifts this restriction.

\paragraph{Promoting code from dynamic to static.}


We can promote groups and control based on inferred \code|@static| annotations.
For example, after inferring the \code|@static(1)| annotation for the group~\code|g|, we can promote it to:
\begin{lstlisting}
static<1> group g { reg.in=10; reg.write_en=1; }
\end{lstlisting}
%
%
%
%
While static control has lower control overhead and enables downstream optimizations,
it incurs two major costs.
We introduce \emph{promotion heuristics} to balance each of these costs.

First, each static island requires
one wrapper interface and one counter register.
This cost is constant for each island, while the benefit of  simpler static control scales with the code size of the island.
Therefore, the compiler introduces a \emph{threshold} parameter that only promotes static islands above a certain code size, in terms of the number of groups and conditional ports.

The second cost affects long-running static islands, which can require large FSM registers and associated comparators.
Some islands reach hundreds of millions of cycles (see \cref{sec:piezo-gen:circt}), so two smaller islands can sometimes be cheaper than one large island.
The compiler accepts a parameter that gives an \emph{upper bound} on the number of cycles that a potential static island can run for, and does not promote any island that would run longer than this upper bound.

We empirically calibrate these parameters' default settings using experience with real programs;
see \cref{sec:piezo-gen:circt}.


\begin{figure}
\centering
\begin{subfigure}[b]{0.45\linewidth}
\includegraphics[width=\linewidth]{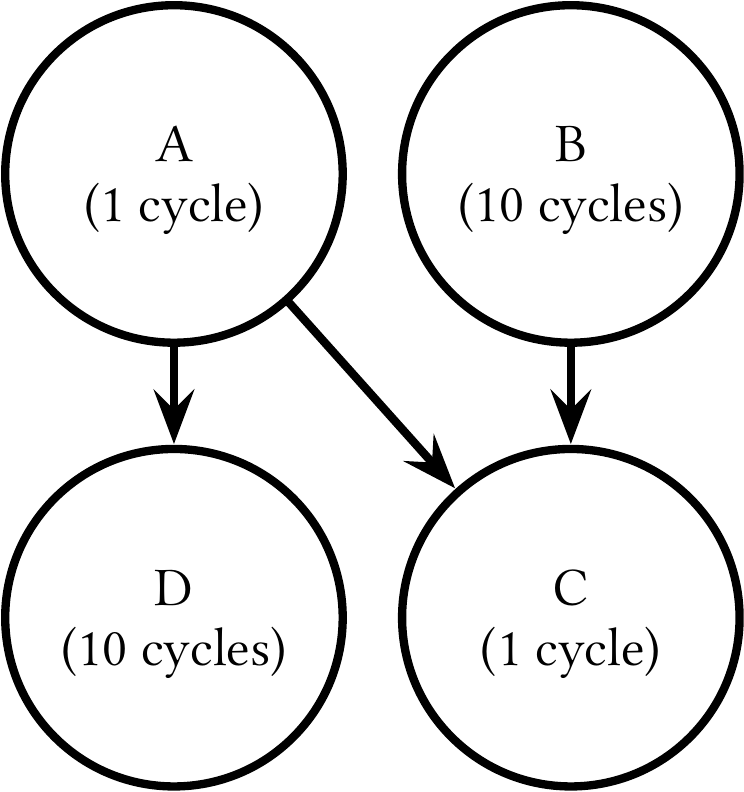}
\caption{Dependency graph.}
\label{fig:compact:deps}
\end{subfigure}
\hfill
\begin{subfigure}[b]{0.5\linewidth}
\begin{lstlisting}
wires {
  // Dummy delay groups
  static<1> group
      del_1 {}
  static<10> group
      del_10 {}
}
static par {
  A; B;
  static seq {
      del_10; C; };
  static seq {
      del_1; D; };
}
\end{lstlisting}
\caption{Compacted schedule.}\label{fig:compact:code}
\end{subfigure}
\caption{Schedule compaction uses data dependencies to generate an \emph{as-soon-as-possible} schedule.}\label{fig:compact}
\end{figure}

\subsection{Schedule Compaction}\label{sec:optimizations:compaction}



\sys{} features a new \emph{schedule compaction} optimization to
maximize parallelism while respecting data dependencies.
Schedule compaction is only feasible in a unified compiler.
In a dynamic IL, the compiler lacks latency information altogether.
In a static IL, the compiler has latency information but is barred from rescheduling code, which could violate timing properties that the program relies on.
Traditional C-based high-level synthesis (HLS) compilers accomplish similar scheduling optimizations,
but by translating between two vastly different representations:
from untimed C to a fully static HDL.
A unified IL, in contrast, can perform this optimization within a single abstraction by exploiting the interaction between static and dynamic code.

Compaction occurs during the transition from dynamic to static code, after \code|@static| inference and as a supplement to standard promotion.
Consider the following \code|seq|:
%
\begin{lstlisting}
@static(22) seq { A; B; C; D; }
\end{lstlisting}
where \Cref{fig:compact:deps} shows the groups' latencies and data dependencies.
If we only perform promotion, it will take $1 + 10 + 1 + 10 = 22$ cycles.

\sys{}'s schedule compaction pass reschedules the group executions
to start as soon as their
dependencies have finished.
Specifically, \code{A} and \code{B} start at cycle 0 because they have no dependencies; \code{C} and \code{D} start on cycle 10 and 1 respectively: the first cycle after their dependencies have finished.
This compacted schedule takes only $11$ cycles.

The optimization extracts a data dependency graph for the children of the \code{seq} and topologically sorts it to produce an as-soon-as-possible schedule.
%
Next, it reconstructs a control program to implement this schedule.
It emits a \code{static par} with one group per thread.
To delay a group's start, it uses an empty delay group, as shown in \cref{fig:compact:code}.
Since all \code{del_n} groups are removed during the collapsing step of compilation (\cref{sec:compilation:collapse}), they incur no overhead.

\subsection{Cell Sharing}\label{section:optimizations:cell-share}

Calyx has a register sharing pass~\cite{calyx} to reduce resource usage.
It uses Calyx's control flow to compute registers' live ranges and remaps them to the same instance when the ranges do not overlap.
\sys{}'s variant is a generalized \emph{cell sharing} pass
that works with arbitrary components instead of just registers.

\sys{} first extends this pass to work with mixed static--dynamic designs and then enhances it to opportunistically exploit static timing information.
%
%
In addition to working uniformly on both static and dynamic code, \sys{}'s cell sharing optimization can
share cells across the static--dynamic boundary: static and dynamic parts of the design can use the same cell.
This is not possible in stratified ILs~\cite{hector,dass} that use separate optimization pipelines for the two interface styles.


\sys{}'s cell sharing pass also improves over sharing in Calyx when it can exploit cycle-level timing in static code.
The original Calyx optimization must
over-approximate live ranges because of Calyx's loose timing semantics.
For example, \code|par| provides no guarantees about the cycle-level timing of its threads (\cref{sec:calyx:static-control}),
so the compiler must conservatively assume that \emph{all} live ranges in one thread may overlap with the live ranges in a different thread.
This prevents Calyx from sharing cells between sibling \code|par| threads.
\sys{}'s enhanced cell sharing optimization exploits timing guarantees (\cref{sec:calyx:static-control}) to compute precise, cycle-level live ranges.
These live ranges are soundly comparable across \code|par| threads and enables sharing between them.
This enhancement
is an example of a \emph{latency-sensitive} optimization from \cref{fig:comp-flow}.

\section{Effects of \sys{} Optimizations}
\label{sec:effects-of-optimizations}


We compare \sys{}'s performance
to Calyx when compiling linear algebra kernels and a packet scheduling engine.

\subsection{Linear Algebra Kernels }
\label{sec:piezo-gen:circt}

CIRCT~\cite{circt-latte} is an MLIR~\cite{mlir} subproject for designing open-source hardware flows.
Calyx is a core dialect within CIRCT and can be generated from C++ or PyTorch programs.
We lower the Polybench benchmarks~\cite{polybench}, written in C++, to Calyx using the CIRCT flow, automatically promote them to use \sys{}'s new abstractions (\cref{sec:optimizations:promotion}), and report the cycle counts and resource usage on an FPGA.

Of the 30 Polybench benchmarks, the CIRCT flow fails to compile 11 due to various limitations in the frontend dialects.
The 19 compilable benchmarks are chiefly dense loop nests,
so large parts of them can be scheduled statically.
However, there is some dynamic behavior, including dynamically-timed integer division and ``triangular'' nested loops (i.e., the inner loop bound depends on the outer loop's index).

\begin{figure}
  \centering
  \includegraphics[width=\linewidth]{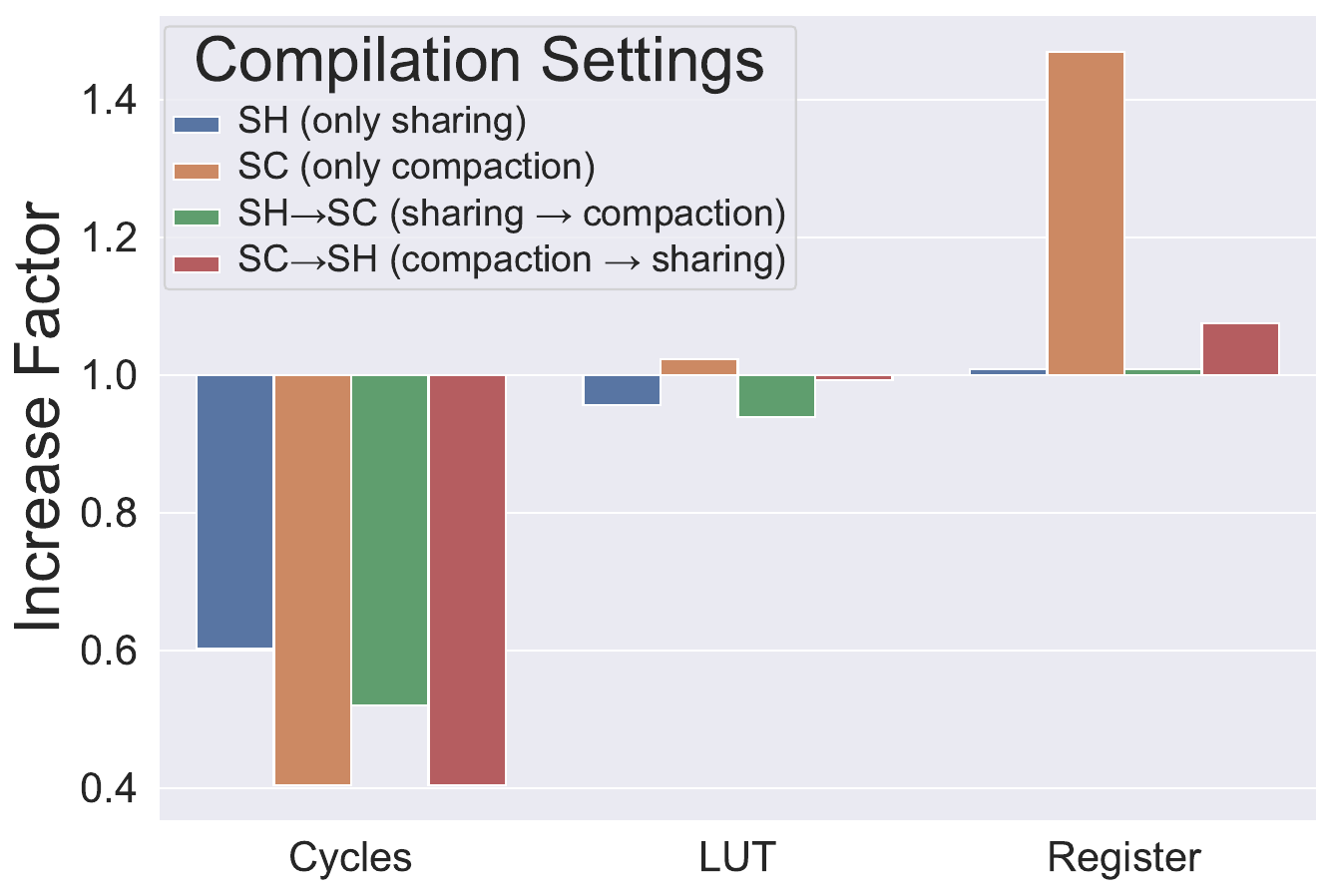}
  \caption{Geometric means across the 19 benchmarks for cycle counts, LUT usage, and register usage. The bars on the graph are normalized to the \textbf{B}aseline, i.e., non-promoted, implementation. Because the geometric means for worst slack (not shown here) among the five configurations are all within 5\% of each other, cycle counts are a fairly accurate measure of runtime.}
  \label{fig:geo-means}
\end{figure}



\paragraph{Configurations.}
To generate \sys{} designs from Calyx, we first perform static promotion (\cref{sec:optimizations:promotion}).
Then, we compile each design with different configurations of the schedule compaction (\textbf{SC}, \cref{sec:optimizations:compaction}) and cell sharing (\textbf{SH}, \cref{section:optimizations:cell-share}) passes.
\begin{enumerate}
\item
\textbf{B}aseline: The standard Calyx compiler, without \sys{} abstractions or promotion.

\item
\textbf{SH}: Static promotion, then cell sharing.

\item
\textbf{SC}: Static promotion, then schedule compaction.

\item
\textbf{SH\textrightarrow SC}: Static promotion, sharing, then compaction.

\item
\textbf{SC\textrightarrow SH}: Static promotion, compaction, then sharing.
\end{enumerate}

\paragraph{Experimental setup.}
We use Verilator v5.006~\cite{verilator} to obtain cycle counts.
Our synthesis flow uses Vivado 2020.2 and targets the Xilinx Alveo U250 board with a period of $7$~ns.
We report post place-and-route resource estimates for lookup tables (LUTs, FPGAs' primary logic resource) and registers.

The geometric mean of the worst timing slack across the 19 benchmarks varies within 5\% between the five configurations.
We therefore believe that $7$~ns is an appropriate clock period for these designs, so measuring cycle counts suffices to reflect actual running time.

We also ran experiments to explore promotion parameters (\cref{sec:optimizations:promotion}).
While they unsurprisingly yield
nonuniform trade-offs between area and latency,
we select default parameters that provide a good balance across benchmarks: 1 for the static island size and 4096 for the cycle-count limit.

\xxx[C]{Add heuristic graphs to appendix?.}

\subsubsection{Schedule Compaction and Cell Sharing}

\Cref{fig:geo-means} compares the configurations normalized to baseline (\textbf{B}).
We report the geometric means for cycle counts, LUTs, and register usage across
the $19$ benchmarks.

\paragraph{Cycle counts.}
Static promotion has significant impact on cycle count.
Running \textbf{SH}, without compaction, can isolate the impact of promotion: a $1.67\times$ geomean speedup over \textbf{B}.
Schedule compaction (\textbf{SC}) improves the cycle count further, yielding a $2.5\times$ geomean speedup over the baseline designs.


\paragraph{LUT usage.}
\textbf{SH} saves LUTs ($0.96\times$) while \textbf{SC} increases them ($1.02\times$), although both are quite similar to \textbf{B}.
\textbf{SH} benefits from \sys{}'s static control interface, while
\textbf{SC} incurs logic overhead to implement its parallelized schedules.


\paragraph{Register usage.}
Sharing hardware resources (\textbf{SH}) performs essentially the same ($1.01\times$) as to \textbf{B}.
This is because~\textbf{B} already has a sharing optimization, and there were not many opportunities to apply the \sys{} extensions to exploit time sensitive sharing (\cref{section:optimizations:cell-share}).
\textbf{SC} incurs a register cost ($1.47\times$) to implement its schedules.

\subsubsection{Phase Ordering}

Schedule compaction and cell sharing are partially in conflict:
the former adds parallelism, while the latter exploits \emph{non-parallel} code to share resources.
They embody a fundamental trade-off between performance and area.
We measure their interaction in either order:
\paragraph{Cycle counts.}
\textbf{SC\textrightarrow SH}
performs identically to
\textbf{SC} alone ($2.5\times$ speedup).
The opposite ordering, \textbf{SH\textrightarrow SC}, is slightly slower ($1.92\times$), but still faster than \textbf{SH} ($1.67\times$).
Sharing impedes some, but not all, opportunities for compaction.

\paragraph{LUT usage.}
\textbf{SH\textrightarrow SC} saves LUTs slightly ($0.94\times$ of \textbf{B}) while
\textbf{SC\textrightarrow SH} performs similarly to \textbf{B} ($0.99\times$).
However, the effects across benchmarks are nonuniform, and the combinations of optimizations can sometimes outperform \textbf{SH} alone.

\paragraph{Register usage.}
Running sharing first (\textbf{SH\textrightarrow SC})
achieves similar register reduction to \textbf{SH} alone
($1.01 \times$ of \textbf{B}).
The reverse ordering (\textbf{SC\textrightarrow SH})
is slightly worse
($1.07\times$) but still significantly better than
\textbf{SC} alone ($1.47\times$).
Running \textbf{SC} first only opportunistically adds parallelism; the designs still have some fundamental sequential behavior that allows sharing.

\subsection{Packet Schedulers}
\label{sec:piezo-gen:pifo}


\begin{figure}
  \centering
    \includegraphics[scale=1.2]{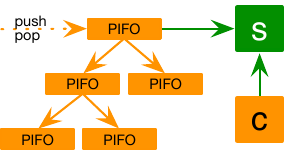}
  \caption{A PIFO tree, a statistics component (\textsf{s}), and a controller (\textsf{c}). Green is static; orange is dynamic.}
\label{fig:pifo}
\end{figure}


We use a second, more domain-specific case study to understand \sys{} optimizations in more detail.
In software-defined networking (SDN)~\cite{sdn},
\emph{programmable packet scheduling} offers flexible policies for allocating bandwidth and ordering packet delivery.
\emph{PIFO trees}~\cite{pifo-siv16, pifo-mohan23} are a flexible mechanism for line-rate packet scheduling.
The packet buffer of a switch consists of a compositional hierarchy of priority queues (PIFOs), each of which implements a policy for scheduling the data held by its children.

We implement a new \sys{}-based generator for PIFO tree packet schedulers as shown in \cref{fig:pifo}.
We \emph{push} incoming packets into the PIFO tree by inserting into a leaf node and adding priority metadata to each parent.
To \emph{pop} the highest-priority packet for forwarding, we query the tree to identify it and update the metadata.
The tree also maintains telemetric data---counts of classes of packets---by reporting to a separate statistics component (\textsf{s}) at each push.
An SDN controller (\textsf{c}) might exploit these statistics to implement adaptive scheduling policies.


Our implementation generates the PIFO tree itself,
which is fundamentally dynamic because of the data-dependent behavior of queues,
and a simple static statistics unit.
While it is not the focus of this case study, we also include a simple dynamic controller to consume the statistics.

\paragraph{Implementation.}


We implement a flexible \sys{} PIFO tree generator in 600 lines of Python.
The generator can produce binary PIFO trees of varying heights, arrangements, and capacities.
It can also implement different scheduling policies
by deciding how packets get assigned to leaves and metadata to internal nodes.
For our experiment, we generate a tree with $5$ PIFOs and overall capacity $10$.
We set up the scheduling parameters to implement a hierarchical round-robin scheduling policy.

We use the generator to synthesize four hardware configurations:
plain Calyx, Calyx promoted to \sys{}, explicitly annotated \sys{}, and annotated, promoted \sys{}.
The second configuration is the result of automatically promoting the first (see \cref{sec:optimizations:promotion}).
The third includes manually inserted \code|static<>| annotations that encode domain-specific insight into the generated hardware's timing.
The fourth configuration is the result of automatically promoting the third.
The generated design is 1,100 lines of \sys{} IL.

\paragraph{Results.}
We generate a workload of 10,000 packets
with randomly interspersed but balanced push and pop events.
We measure
the LUT count, register count, and cycles per push (C/push) for each design.
The best values are in bold.

\noindent
\begin{minipage}[c]{\linewidth}
\centering
\vspace{5pt}
\begin{tabular}{crrr}
\toprule
Configuration & LUTs & Registers & C/push \\
\midrule
Calyx & 1547 & 393 & 143.25 \\ 
Promoted to \sys{} & 1610 & 391 & 139.25 \\ 
Annotated \sys{} & 1556 & 385 & 140.25 \\ 
Annotated, Promoted \sys{} & \textbf{1544} & \textbf{381} & \textbf{137.25} \\ 
\bottomrule
\end{tabular}
\vspace{5pt}
\end{minipage}
The resource usage of the three designs is similar: the PIFO tree (which is always dynamic) is the dominant component in all three designs, and the statistics component (which is dynamic in Calyx and static in \sys{}) is small.

The promoted \sys{} implementation improves on the original Calyx implementation's C/push measure because the promotion pass exploits small opportunities for static promotion in all components, including components that are understood to be dynamic.
This comes at the cost of resource usage: promotion to \sys{} also triggers schedule compaction (\cref{sec:optimizations:compaction}), and compaction costs LUTs.
The manually annotated \sys{} implementation also improves on the baseline's C/push measure---domain knowledge lets the human guide the compiler---but \emph{without} suffering a LUT cost.
The annotated, promoted \sys{} implementation performs the best of all.
Its C/push measure is better than the manually annotated \sys{} implementation because, as before, the promotion pass exploits opportunities for promotion that are not clear to a human.
Critically, its LUT count does \emph{not} suffer compared to annotated \sys{}: compaction only runs on promoted static islands, not user-defined static components.

\section{Systolic Arrays}
\label{sec:piezo-gen:systolic}

\begin{figure}
  \centering
  \includegraphics[width=\linewidth]{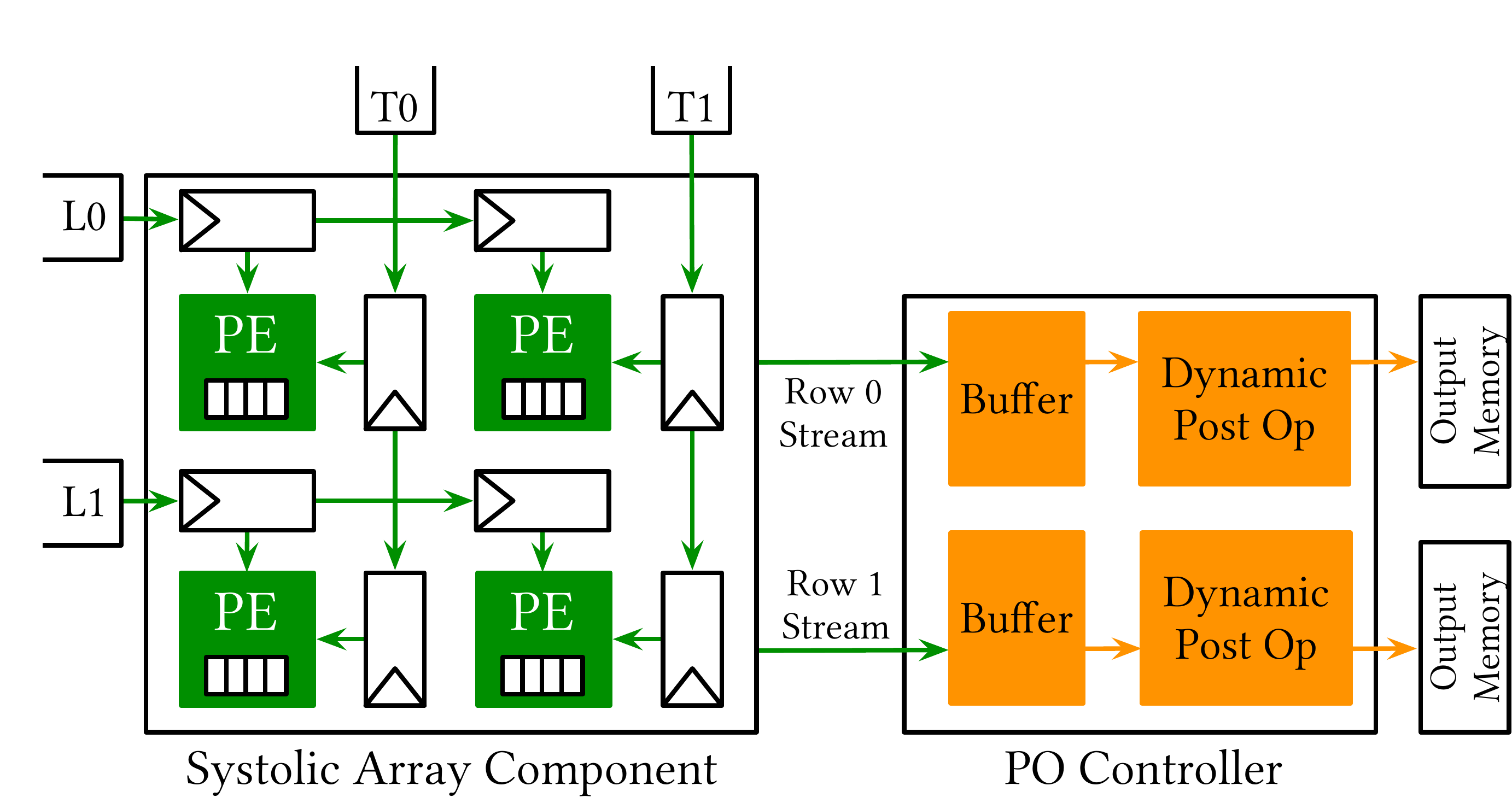}
  \caption{Our $2 \times 2$ systolic array with a \textit{dynamic} post op.
  The buffers in the post op controller are not necessary for static post ops.
  Green is static and orange is dynamic.
  Input memories (\code{L0}, \code{L1}, \code{T0}, \code{T1}) may have non-fixed length.
  }
  \label{fig:systolic-design}
\end{figure}

Systolic arrays~\cite{kung:systolic} are a class of architecture commonly used in machine learning~\cite{brainwave,tpu} built from interconnected processing elements (PEs).
PEs perform simple computations and communicate with other PEs in a simple, regular manner.
We redesign an existing systolic array generator that targets Calyx to use \sys{} abstractions and demonstrate how it enables efficient composition and incremental adoption.

\subsection{Systolic Arrays in \sys{}}

Calyx has an existing systolic array generator that produces hardware to multiple fixed-size matrices.
The interface of the generated systolic array accepts rows and columns of input matrices $A$ and $B$ in parallel using an output-stationary dataflow.
Each PE performs a multiply-accumulate operation and forwards its operands.

This case study addresses three main limitations: 
\begin{itemize}
\item
Calyx's dynamic interfaces between PEs make it challenging to pipeline computations, hindering performance.
\sys{} enables efficient pipeline execution using static interfaces.


\item
A purely static implementation can only efficiently support fixed-sized matrices.
\sys{}'s unified approach makes it possible to support flexible matrix sizes while maintaining efficient pipelined execution.


\item
Systolic arrays often have fused \emph{post operations} that apply elementwise functions to the product matrix.
We show how \sys{}'s mixed interfaces support various post post ops and optimize the composed design across the static--dynamic boundary.

\end{itemize}

\paragraph{Pipelining processing elements.}
\label{sec:systolic:pipelining}


Calyx's systolic array generator decouples the logic for the PE from the systolic array itself to modularize code generation.
This means that the systolic array must communicate with its PEs through dynamic interfaces.
Because there are no timing guarantees, the generator does not pipeline the PEs and instead uses sequential multipliers.
Extending the generator to a dynamically pipelined design would add unnecessary overhead; we would need queues to buffer values between PEs.

Instead, \sys{} abstractions let the systolic array communicate with its PEs using efficient static interfaces that facilitate pipelining.
Besides removing the overhead from dynamic interfaces, this also simplifies the logic of the systolic array fabric, which is in charge of data movement.
Because we have a pipeline with initiation interval of $1$, the fabric can unconditionally move data every cycle and guarantee the right value will be read.

\paragraph{Fixed contraction dimension.}
While static interfaces allow for efficient, pipelined execution, they can limit computational flexibility.
For example, an output-stationary matrix-multiply systolic array
should be able to multiply matrices of sizes $i \times k$ and $k \times j$ for any value of $k$.
However, this requires dynamic control flow: the computation needs to repeat $k$ times where $k$ is a runtime value.
\sys{} abstractions support this with ease: we use a \code|while| loop to execute the systolic array's logic $k$ times.
Furthermore, because the control program in the loop body is purely static, \sys{}'s special handling ensures that the body executes every cycle (\cref{sec:compilation:groups}).

\paragraph{Supporting fused post operations.}

A common optimization in machine learning frameworks~\cite{intelDNN}
\emph{fuses} matrix multiplication with elementwise \emph{post operations}, such as nonlinearities, to avoid writing the intermediate matrix back to memory.
These post operations can be either fundamentally static or dynamic.
Our goal is to decouple the implementation of post operations from the systolic array:
to keep the code generation modular without sacrificing efficient interfaces.
We implement two post operators (POs):
(1) a static ReLU operation, \code|x > 0 ? x : 0|, and
(2) a dynamic leaky ReLU~\cite{rectifiers} operation, \code|x > 0 ? x : 0.01*x|.
The latter is dynamic because the true branch can directly forward the output while the false branch requires a multiplication.

\Cref{fig:systolic-design} overviews the architecture.
We instantiate the systolic array and PO components for the number of rows in the resulting matrix.
If the PO is dynamic, the \emph{PO controller} instantiates buffers to queue the output stream but elides them for static POs.
The interface between the systolic array and PO is pipelined: a row's PO starts its computation as soon as an output is available.
Most of the code---the systolic array, the controller, the PEs---is reused regardless of the PO's interface; \sys{}'s unified abstractions enable this reuse.

\subsection{Evaluation}\label{sec:results:systolic}

\begin{figure*}
  \centering
  \begin{subfigure}[b]{0.32\linewidth}
    \centering
    \includegraphics[width=\linewidth]{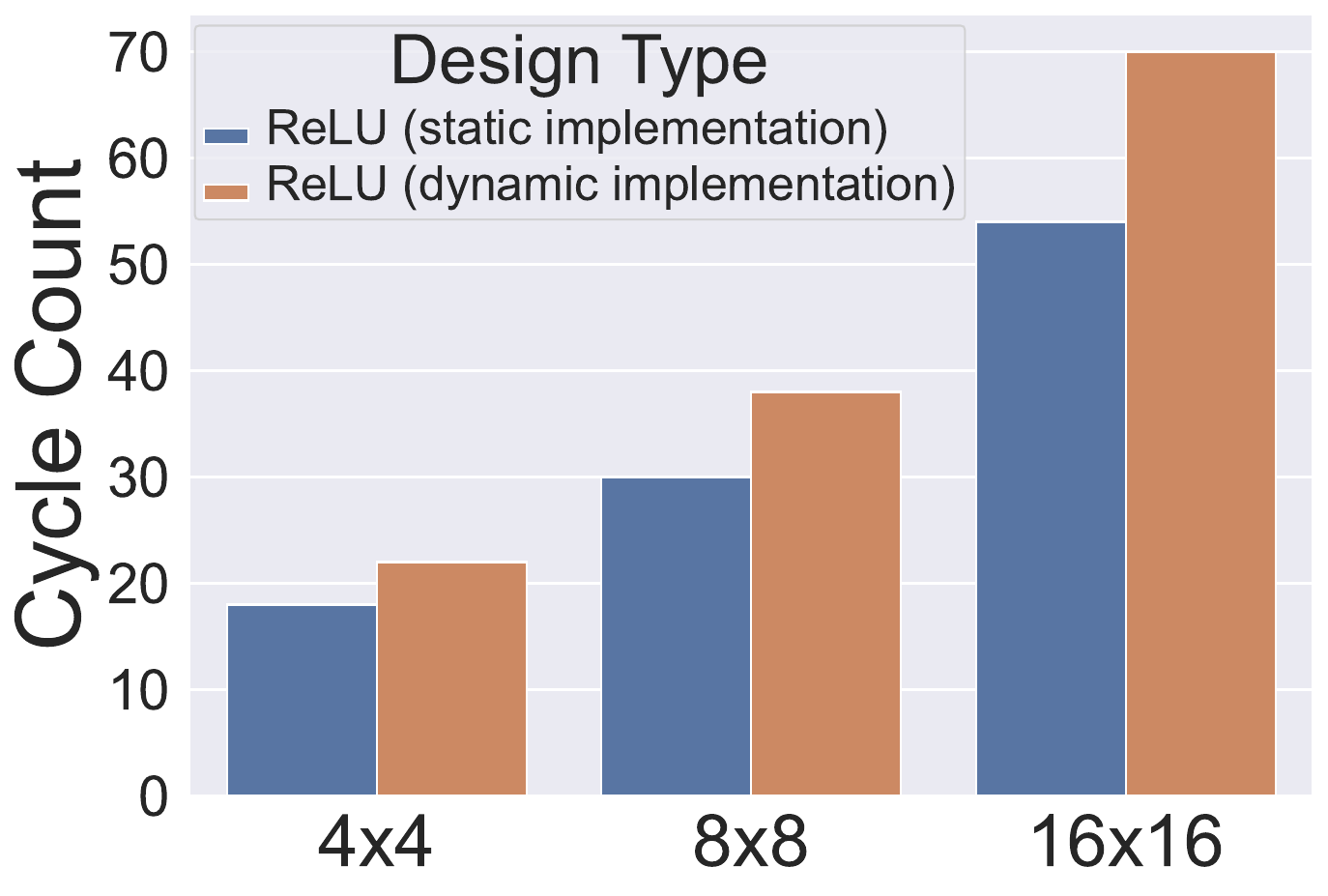}
    \caption{Cycle counts.}
    \label{fig:relu-cycles}
  \end{subfigure}
  \begin{subfigure}[b]{0.32\linewidth}
    \centering
    \includegraphics[width=\linewidth]{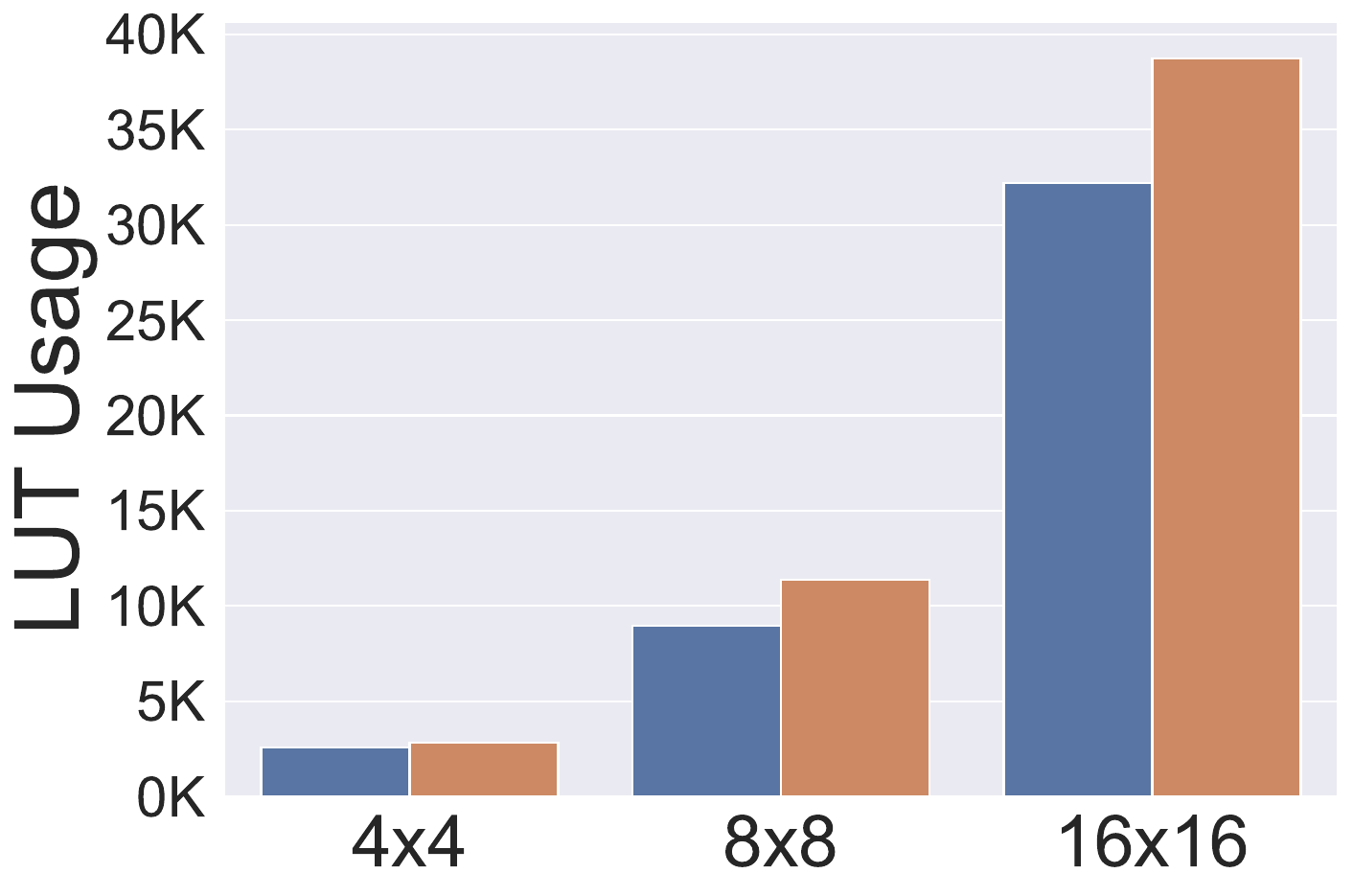}
    \caption{LUT usage.}
    \label{fig:relu-lut}
  \end{subfigure}
  \begin{subfigure}[b]{0.32\linewidth}
    \centering
    \includegraphics[width=\linewidth]{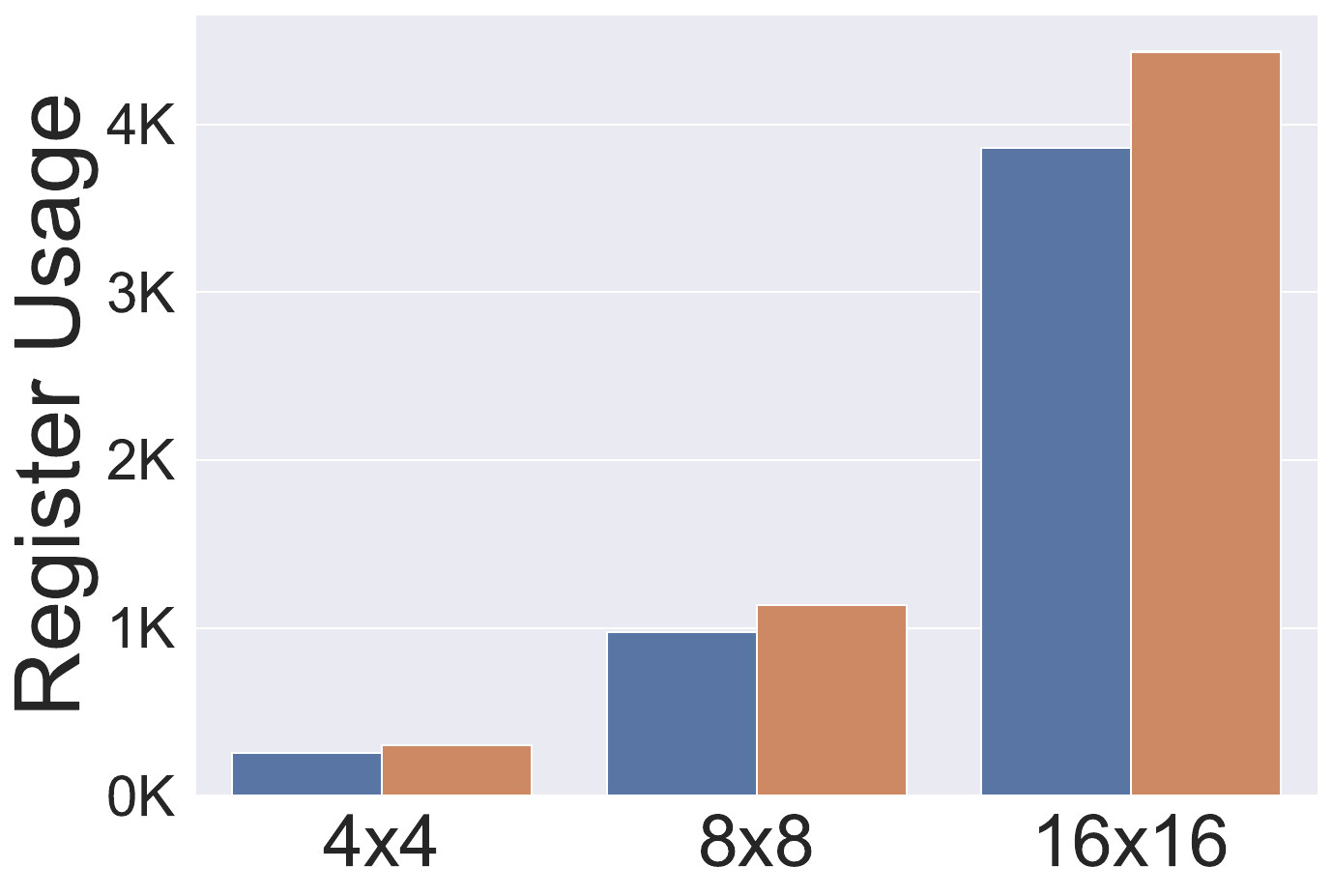}
    \caption{Register usage.}
    \label{fig:relu-reg}
  \end{subfigure}
  \caption{Performance and FPGA resource utilization of two implementations of a fused matrix-multiply--ReLU kernel on \sys{}-compiled systolic arrays.
  We compare static and dynamic interfaces for the ReLU unit.}
  \label{fig:systolic-graphs}
\end{figure*}

Our evaluation seeks to answer the following questions:
\begin{itemize}
\item
Does the pipelined \sys{}-generated systolic arrays outperform the existing Calyx-generated designs?

\item
Can \sys{} implement a runtime-configurable contraction dimension for systolic arrays with low overhead?

\item
Do cross-boundary optimizations let \sys{} eliminate overheads when the systolic array is coupled with a static post operation?
\end{itemize}

\paragraph{Effect of pipelining.}
For the $16 \times 16$ design, the pipelined implementation in \sys{} achieves a max frequency of $270$ MHz and performs the computation in $52$ cycles in comparison to the original design's $250$ MHz and $248$ cycles.
The latency improvement is from the pipelined execution and the frequency improvement from simplified control logic.

\paragraph{Configurable matrix dimensions.}
We compare systolic arrays with \emph{flexible} and \emph{fixed} matrix size support.
The flexible design takes $1$ extra cycle to finish, uses $8\%$ more LUTs (for logic to check the loop iteration bound), and uses the same number of registers.
The flexible design pays some overhead to gain dynamic functionality, while the fixed design is fully static, thereby eliminating dynamic overhead: \sys{} expresses both with minimal code changes.


\paragraph{Overhead of dynamic post operations.}
We perform a synthetic experiment to quantify overhead of a dynamic interface between the systolic array and the PO:
we use the simple ReLU post operation in its default, static form
and compare it against a version that artificially wraps it in a dynamic interface.
Since the computation is the same, the only difference is the interface.
\Cref{fig:relu-cycles,fig:relu-lut,fig:relu-reg} report the cycle counts, LUTs, and register usage of the resulting designs.
In addition to a higher cycle count, the dynamic implementation also has higher LUT and register usage, stemming from the extra control logic and buffers respectively.

We also implemented a truly dynamic post operator,
leaky ReLU.
We omit its measurements here because it conflates the costs of the operation and static--dynamic interaction.

\section{Related Work}

\sys{} builds on a rich body of prior work on compilers for accelerator design languages (ADLs): high-level programming models for designing computational hardware.
However, these compilers tend to prioritize either static or dynamic interfaces in the hardware they generate---or, when they combine both strategies, to disallow fluid transitions between the two styles.

Traditional C-based high-level synthesis (HLS) compilers \cite{autopilot,bambu,legup,intelhls,catapulthls,stratushls}
intermix static and dynamic-latency operations, such as dividers.
They do so using software ILs like LLVM~\cite{llvm}, which ties them to C-like, sequential computational models.
Critically, traditional HLS tools are monolithic:
they do not expose consistent intermediate representations that support
modular pass development,
decoupled frontends and backends,
and layered correctness arguments.
\sys{} contributes a stable IL that includes both software- and hardware-like abstractions and thus supports modular passes that address both static and dynamic control.


The most closely related compilers seek to combine aspects of static and dynamic control~\cite{hector, dass}.
DASS~\cite{dass} is the first HLS compiler we are aware of to specifically balance static and dynamic scheduling within the same program.
In DASS, either the user~\cite{dass} or some heuristic~\cite{dass-islands} identifies parts of the high-level design that would benefit from static scheduling.
Compilation proceeds in two phases:
DASS first compiles all the static islands, and then it uses a second, dynamic, approach to schedule the rest of the program while treating the pre-compiled islands as opaque operators.
In contrast, \sys{}'s unified IL can treat static portions of the program transparently and optimize them in the same framework as dynamic code.
Szafarczyk et al.~\cite{elastic-sycl-hls} provide the opposite approach to DASS: it finds sections of programs that are amenable to dynamic scheduling in a previously statically-scheduled program.
The dynamic sections are decoupled from the static parts and compiled into processing elements that communicate over latency-insensitive channels.
Hector~\cite{hector} is a dialect of MLIR~\cite{mlir} that
supports three scheduling styles: pipeline, static, and dynamic.
Each style corresponds to a different Hector component type,
uses a different a syntax and semantics,
and uses a different lowering strategy.
In contrast to these, \sys{} provides a unified IL in which either the frontend, or a compiler heuristic (\cref{sec:optimizations:promotion}), can easily covert dynamic programs to static and vice-versa, and lowers them using a single compilation pipeline.
This lets \sys{} reuse optimizations between the two modes and even optimize across the boundary between dynamic and static code.




Other ILs for ADL compilers also give passes control over scheduling, but focus on either static~\cite{ahir,synasm,suifhls} or dynamic interfaces~\cite{dynamic-schedule-hls,muir}.
%
In particular, HIR~\cite{hir} is an MLIR-based IL that describes schedules using
\textit{time variables} that describe the clock cycles on when each value in a design is available.
Filament~\cite{filament}, like HIR, explicitly dictates the cycle-level schedule of hardware operations, but it encodes these time intervals into a type system.
\sys{}'s relative timing guards (\cref{sec:calyx:static-groups}) work similarly and describe the cycle-level schedule for assignments.
However, \sys{}'s timing guards are relative to the start of each group's execution.
This \emph{relative} timing limits the scope of static schedules and enables flexible composition with dynamic groups, scalable reasoning, and efficient lowering (\cref{sec:compilation:collapse}).
Finally, unlike both systems, \sys{} supports both static and dynamic interfaces.

%

\section{Conclusion}

Latency-sensitive hardware refines the semantics of latency-insensitive hardware.
Every practical accelerator compiler must combine the two styles,
and this correspondence is the foundation for combining them soundly.

\bibliographystyle{plain}
\bibliography{./bib/venues,./bib/papers}

\end{document}